\documentclass[a4paper]{jpconf}
\usepackage{graphicx}
\begin{document}
\title{Isotopic fission fragment distributions as a deep probe 
to fusion-fission dynamics 
}

\author{F. Farget$^1$, M. Caama\~no$^2$, O. Delaune$^1$, O. B. Tarasov$^3$,  X. Derkx$^{1,}$\footnote[11]{Present address:
School of Engineering, University of the West of Scotland, UK}, K.-H. Schmidt$^{1,}$\footnote[12]{On leave from: GSI, Planckstrasse 1, 64291 Darmstadt, Germany}, 
A. M. Amthor$^{1, 3}$, L. Audouin$^4$, C.-O. Bacri$^4$, G.~Barreau$^7$,
B. Bastin$^1$, D. Bazin$^3$, B. Blank$^7$, J. Benlliure$^2$, L.~Cac\'eres$^1$, E. Casarejos$^6$, A. Chibihi$^1$,
B.~Fern\`andez-Dominguez$^2$, L.~Gaudefroy$^5$,
C. Golabek${^1}$,  S. Gr\'evy$^{1,}$\footnote[13]{Present address: CENBG, UMR 5797 CNRS/IN2P3, Universit\'e Bordeaux 1, F-33175 Gradignan, France}, B. Jurado$^7$,
O. Kamalou$^1$, A.~Lemasson$^1$, S.~Lukyanov$^9$,
W. Mittig$^{3, 8}$, D.~J.~Morrissey$^{3, 10}$, 
A.~Navin$^1$,  J. Pereira$^3$, L.~Perrot$^6$, 
M. Rejmund$^1$, T. Roger$^1$,  M.-G.~Saint-Laurent$^1$, H.~Savajols$^1$,  C.~Schmitt$^1$, B. M. Sherill$^{3, 8}$, C.~Stodel$^1$, J. Taieb$^5$, 
J.-C.~Thomas$^1$, 
A.~C.~Villari$^{1,}$\footnote[14]{Present address: Pantechnik S.A., 13, rue de la R\'esistance, 14400, Bayeux, France}
}

\address{$^1$GANIL, , CEA/DSM-CNRS/IN2P3, BP 55027, F-14076 Caen cedex 5, France}
\address{$^2$USC, Univ. de Santiago de Compostela, E-15706 Santiago de Compostela, Spain}
\address{$^3$NSCL, Michigan State University, East Lansing, MI 48824, USA}
%\address{$^4$GSI, Planckstrasse 1, 64291 Darmstadt, Germany}
\address{$^4$Institut de Physique Nucl\'eaire, CNRS/IN2P3, F-91406 Orsay, France}
\address{$^5$CEA, DAM, DIF, F-91297 Arpajon, France}
\address{$^6$University of Vigo, E-36310 Vigo, Spain}
\address{$^7$CENBG, UMR 5797 CNRS/IN2P3, Universit\'e Bordeaux 1, F-33175 Gradignan, France}
\address{$^8$Dep. of Physics and Astronomy, Michigan State University, East 
Lansing, MI 48824, USA}
\address{$^9$FLNR, JINR, 141980 Dubna, Moscow region, Russian Federation}
\address{$^{10}$~Dep. of Chemistry, Michigan State University, East Lansing, MI 48824, 
USA}

\ead{fanny.farget@ganil.fr}

\begin{abstract}
During the fission process, the atomic nucleus deforms and elongates up to the two fragments inception and their final separation at the scission deformation. The evolution of the nucleus energy with deformation defines a potential energy landscape in the multi-dimensional deformation space. It is determined by the macroscopic properties of the nucleus, and is also strongly influenced by the single-particle structure of the nucleus, which modifies the macroscopic energy minima. The fission fragment distribution is a direct consequence of the deformation path the nucleus has encountered, and therefore is the most genuine experimental observation of the potential energy landscape of the deforming nucleus.  Very asymmetric fusion-fission reactions at energy close to the Coulomb barrier,  produce well-defined conditions of the compound nucleus formation, where processes such as quasi-fission, pre-equilibrium emission and incomplete fusion are negligible. In the same time, the excitation energy is sufficient to reduce significantly structural effects, and mostly the macroscopic part of the potential is  responsible for the formation of the fission fragments. We use inverse kinematics combined with a spectrometer to select and identify the fission fragments produced in $^{238}$U+$^{12}$C at a bombarding energy close to and well-above the Coulomb barrier. For the first time, the isotopic yields are measured over the complete atomic-number distribution, between Z=30 and Z=63. In the experimental set-up, it is also possible to identify transfer-induced reactions, which lead to low-energy fission where the nuclear shell structure shows a strong influence on the fission-fragment distributions. The resulting set of data gives the possibility to observe the fission fragment properties over a wide range of excitation energy, and they reveal the vanishing of the shell effects in the potential energy of the fissioning nucleus, as well as the influence of fission dynamics.
\end{abstract}

\section{Introduction}
Fragment mass distributions produced in low excitation-energy fission of trans-uranium actinides are known for a long time to present in general a double-humped structure, the fragments being divided in two distinct groups, a heavy-fragment group and a light-fragment one~\cite{Unik}.  With increasing mass of the fissioning nucleus, the heavy-fragment group shows a constant average mass, close to the value A=140, whereas the  average mass of the light-fragment group increases. This double-humped structure has been understood as a signature for shell-structure influence in the potential energy describing the deforming nucleus from the ground state to the scission point deformation. The constant position of the heavy fragment distribution lead to the comprehension that the shell influence is in fact the single-particle structure inside the nascent fragments, and not in the fissioning nucleus. The most spread model for the description of the fission fragment distribution is based on the macro-microscopic description of the fissioning nucleus potential energy~\cite{Wilkins, MoellerPRC61,Sida, GEF}, whereas recent approaches based on a fully microscopic description have been attempted~\cite{Goutte}.  The main idea in the macro-microscopic description of fission is to calculate the nucleus potential energy surfaces in a multi-dimensional space taking into account the different possible deformations of the nucleus. In the model of Wilkins {\it et al}~\cite{Wilkins}, only the scission point is considered to impact on the scission configuration, whereas in the M\"{o}ller and Randrup prescription~\cite{MoellerPRC61}, the evolution of the fission barrier with asymmetry is the main issue to determine the final scission split. These differences are correlated to different underlying suppositions on the physics that drive the deforming nucleus, either a purely statistical process driven only by the available phase space in the final states, either a more dynamically evolutive process, where each step in deformation has an impact on the following one, and may influence the final split. 
In any case, the description of the shell structure at large deformation is a key issue to estimate its influence on the gross properties of the potential energy surface. It is commonly admitted that neutron shell gaps appearing for spherical shape and for scission deformation play the most important role in the modification of the potential energy surface. Because the liquid-drop potential energy surface shows minimum for large deformations of both fragments, it is the neutron deformed shell gap ($N \simeq 90$)  that has the most influence on the final yields, whereas the spherical ($N = 82$) one has smaller impact, despite its larger amplitude.  This is why in general, the most produced mass is around 140, whereas  $^{132}$Sn is not among the most produced isotopes. The influence of shell structure is in general computed with the Strutinsky method, and different parameters enter into account; as the potential chosen, and its development with deformation,  the integral region for the averaging of the nuclear level distances, etc. Consequently, the shell gap predictions at scission deformation are difficult to predict firmly, and are difficultly confirmed by experiment. Indeed, there exists no information on the single-particle structure at such important deformation, and in general, only mass distributions of fission fragments are measured, which result from the sum of proton and neutron orbital filling. Different shell gaps are therefore predicted or considered to appear at scission deformation, from N=86~\cite{Wilkins} to N=92~\cite{SchmidtEPL83}. Proton shell gaps are usually not considered, as for large deformation they are predicted to be of much smaller influence on the potential energy than neutron shell; 1 MeV compared to 4 MeV~\cite{Wilkins}.

However, an experiment performed in GSI~\cite{schmidt2000}, in which fission was induced in electromagnetic interaction between a series of actinides and pre-actinides, produced in the fragmentation of a primary $^{238}$U beam, and a lead target, gave access for the first time to the atomic number distribution of the fission fragments over an unprecedented large range of fissioning nuclei. These results of a new kind revealed that when asymmetric, the heavy group of the fission fragment distribution is centred around a constant atomic number, $Z=54$, independently of the fissioning system. This raised the question of the influence of a possible deformed shell gap at the scission deformation for this number of protons~\cite{benlliureEPJ13}, which is not predicted by any calculation~\cite{ragnarsson}. In order to deepen our understanding on the role of neutron and proton shell gaps in the evolution of the potential energy surface, it is clear that more precise data are needed on the proton and neutron distribution in fission fragments. 

In heavy-ion induced fission reactions, the compound nucleus is produced with some excitation energy. The evolution of the potential energy with the different collective parameters is not sufficient to describe the fission-fragment distribution. In addition to its static description, the evolution of the potential energy and the shell structure effects with the excitation energy has an impact on the fission rate as well as on the fission-fragment properties. The time needed for the compound nucleus to deform from its initial shape to the saddle point rules the excitation energy released by particle evaporation before the saddle point, and consequently also influences the fission-fragment distribution. The different time scales find their origin in the propensity the nucleus has to release its intrinsic excitation energy in collective excitation, a process which is governed by nuclear viscosity. 

In this work we report on the results of different experiments based on the inverse kinematics technique, coupled to two spectrometers of GANIL~\cite{vamos,LISE}, which allow for measuring with a good resolution the isotopic (mass A and atomic number Z) distributions of fission-fragments over the complete atomic-number distribution. The different reactions use a $^{238}$U beam at different impinging energies on light targets of $^{12}$C and $^{9}$Be. The extreme asymmetric reactions lead to the formation of compound nuclei produced in the fusion of the beam and the target, with negligible contribution of quasi-fission~\cite{donadille}. In addition to fusion reactions, the experimental set-up allows to disentangle the transfer-induced fission. In transfer reactions, the excitation energy transferred is of the order of 10 MeV, which means only few MeV above the fission barrier, and therefore influence of the single-particle structure in neutron and proton numbers of the fission fragments are expected to appear. Using impinging beam energies ranging from 6 A MeV to 24 A MeV, the compound nuclei are produced in different regimes of excitation energy. The comparison of the fission-fragment properties produced in high-energy fission and in transfer-induced fission is used to look into the time scale of the deformation process.

 \section{Fission experiment using inverse kinematics at Coulomb energy}

\begin{figure}[h]
%\begin{minipage}{14pc}
\includegraphics[width=30pc]{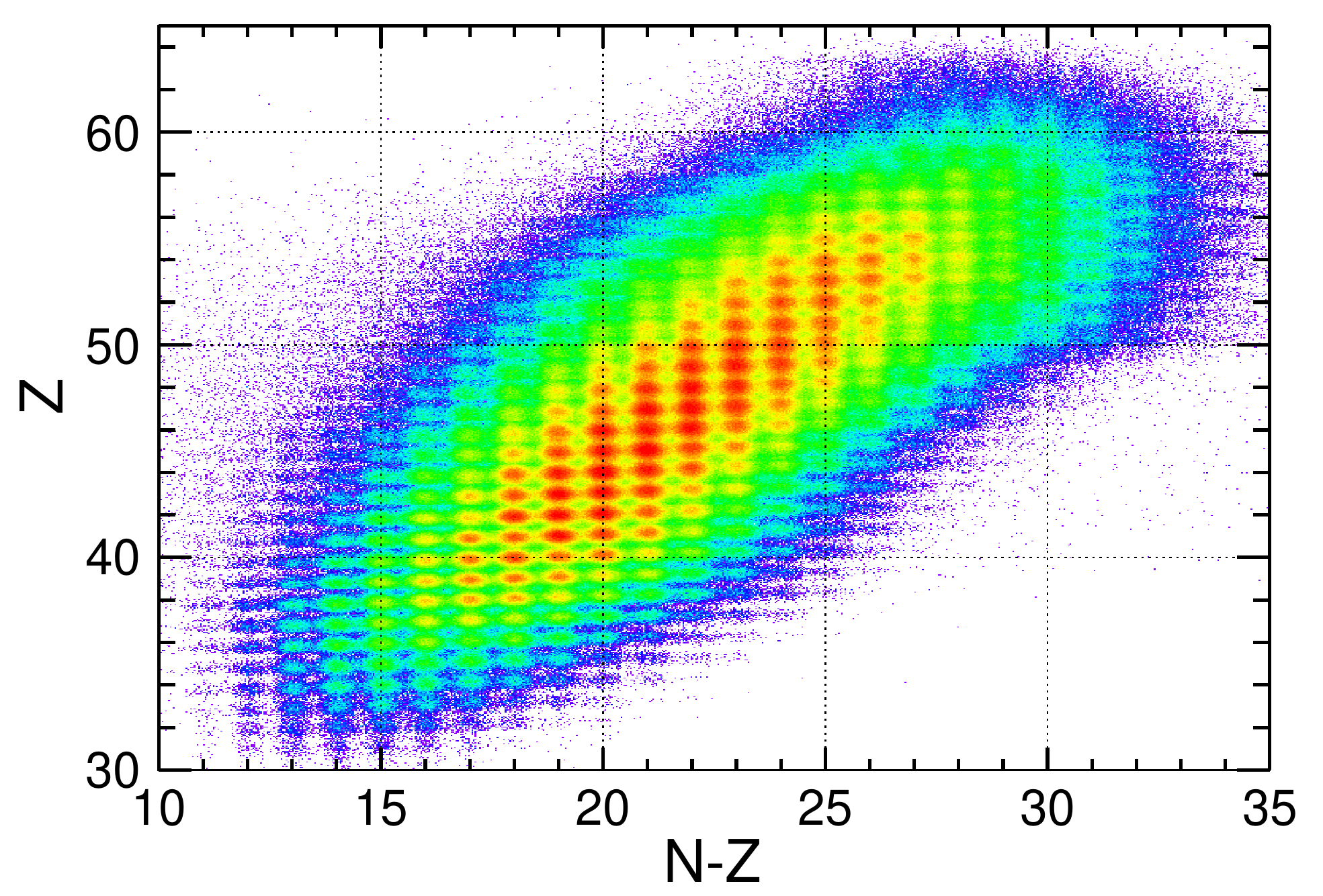}
\caption{\label{Z_N-Z}Identification matrix (Z:N-Z) of the fragments produced in the $^{238}$U+$^{12}$C reaction at 6.1 A MeV.}
%\end{minipage}\hspace{2pc}%
\end{figure}

The inverse kinematics technique for studying fission-fragment properties has been developed at GSI at relativistic energies. At this energy, fission is induced by spallation or fragmentation~\cite{enqvist, bernas} or electromagnetic interaction of secondary beams~\cite{schmidt2000}. At GANIL, the Uranium beam energy is limited from 5 to 25 A MeV, and therefore the nuclear reactions accessible are nucleon transfer or fusion. Using the first cyclotron of GANIL, the $^{238}$U beam is accelerated up to 6.1 A MeV to impinge a $^{12}$C  target, 100$\mu g/cm^2$ thick, wich correspond to an energy in the centre of mass 10\% above the Coulomb barrier. In such asymmetric entrance channel, the formation of a compound nucleus corresponding to the complete fusion of the target and the beam nuclei is the most probable process. Multi-nucleon transfer reactions arise to a total of about 10\% of the total reaction cross section~\cite{Biswas}, and produce different actinides from U to Cm~\cite{Derkx09}, associated to an energy  dissipation in the collision being ruled by the optimal $Q$ value~\cite{Karp}. Excitation-energy distributions are typically centred around 10 MeV, with a standard deviation of typically few MeV. In the case of fusion reactions, the compound nucleus is produced with an excitation energy of 45 MeV, where shell effects are expected to vanish. The VAMOS spectrometer rotated to 20$^\circ$ is used to identify the fission fragments in atomic and mass number $Z$ and $A$, as well as in ionic charge state $q$.  
Depending on their emission direction in the fissioning-nucleus reference frame, the fission fragments are produced with a kinetic energy in the laboratory reference frame ranging from 2 A MeV (backward direction to the beam) to 8~A~MeV (forward direction to the beam) approximatively, with an aperture cone limited to 25$^\circ$. At this kinetic energy, each isotope presents a broad ionic charge state distribution extending over more than 10 different values. Despite of this broad ionic charge state distribution, the energy loss in an ionisation chamber placed at the focal plane of the spectrometer allows for an identification of the atomic number of the ions, with a resolution of $\Delta Z/Z \simeq 1.5 \%$, which provides a good separation of the different elements up to $Z=63$ (Eu). The kinetic properties of the ions, such as their total energy and velocity are measured within a silicon wall and a time-of-flight detector placed behind and before the ionisation chamber, respectively~\cite{Pulli}. The magnetic rigidity of the fragments is deduced from their position and angle at the focal plane, and a reconstruction procedure that takes into account for the aberrations induced by the large angular aperture of VAMOS~\cite{Rejmund}. A mass resolution of $\Delta A/A \simeq 0.6 \%$ is obtained. A detailed description of the identification and calibration procedure is given in~\cite{Caamano09}. Figure~\ref{Z_N-Z} shows the identification matrix ($Z,N-Z$) for the ensemble of the fragments produced in the reaction.

\begin{figure}[h]
%\begin{minipage}{14pc}
\includegraphics[width=30pc]{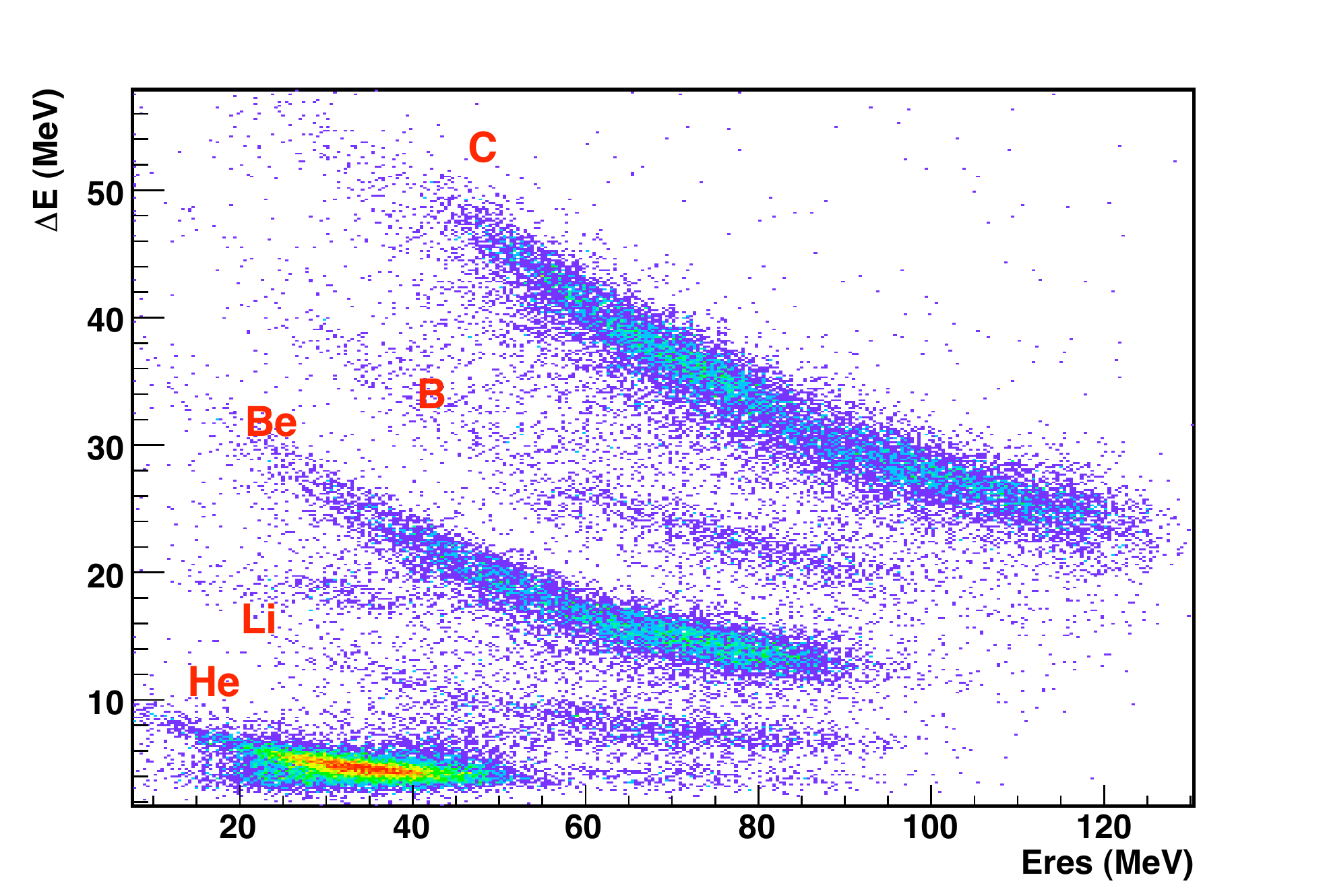}
\caption{\label{spider_dEE}Identification matrix of the target recoils produced in the $^{238}$U+$^{12}$C reaction at 6.1 A MeV. Energy loss of the recoils is plotted against their residual energy. The different proton-transfer channels are identified. The matrix is produced  with a fission fragment detected in VAMOS in coincidence. }
%\end{minipage}\hspace{2pc}%
\end{figure}

\begin{figure}[h]
\begin{minipage}{14pc}
\includegraphics[width=14pc]{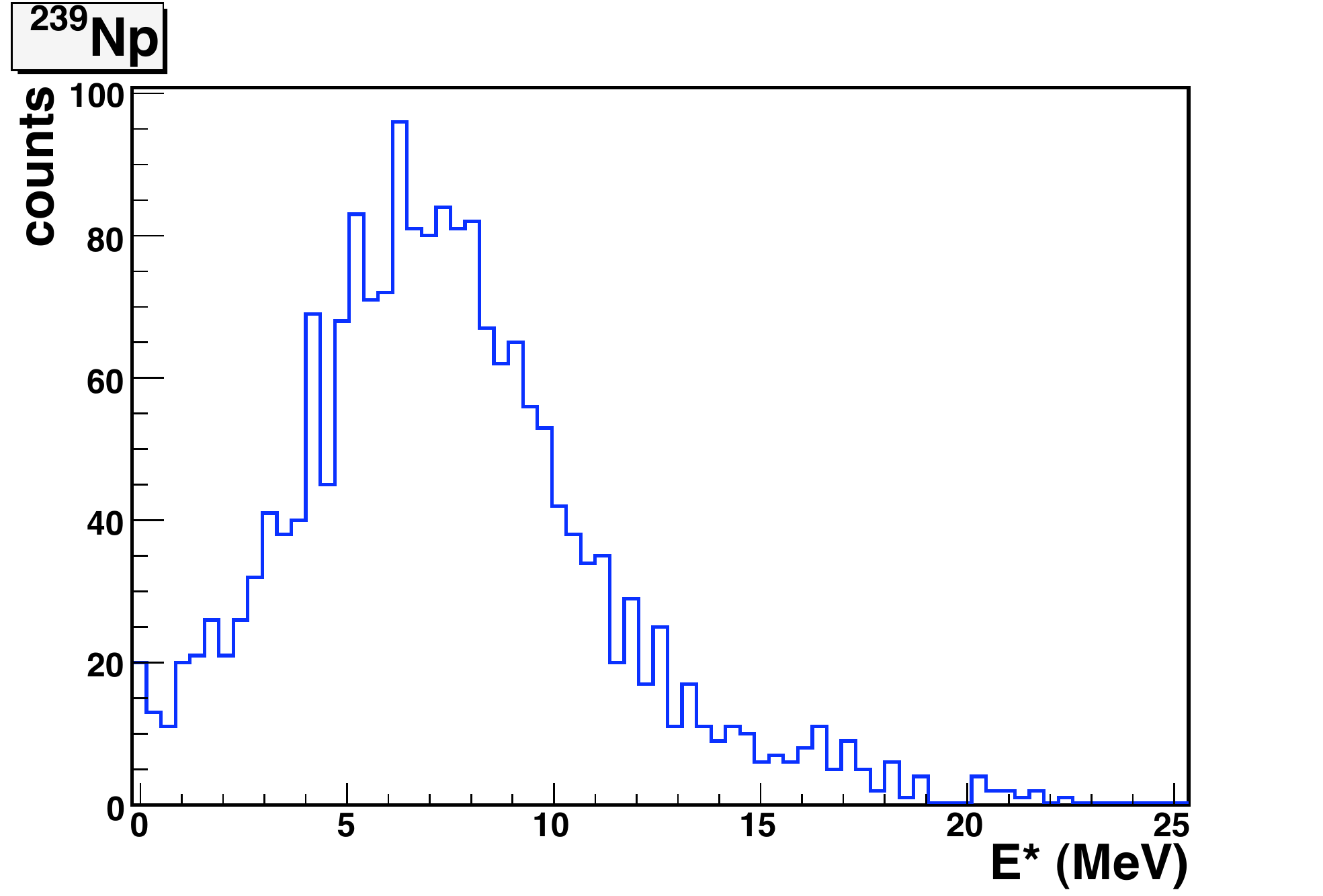}
%\caption{\label{spider_eex_Be}Excitation energy spectrum associated to the B recoils.  }
\end{minipage}\hspace{2pc}%
\begin{minipage}{14pc}
\includegraphics[width=14pc]{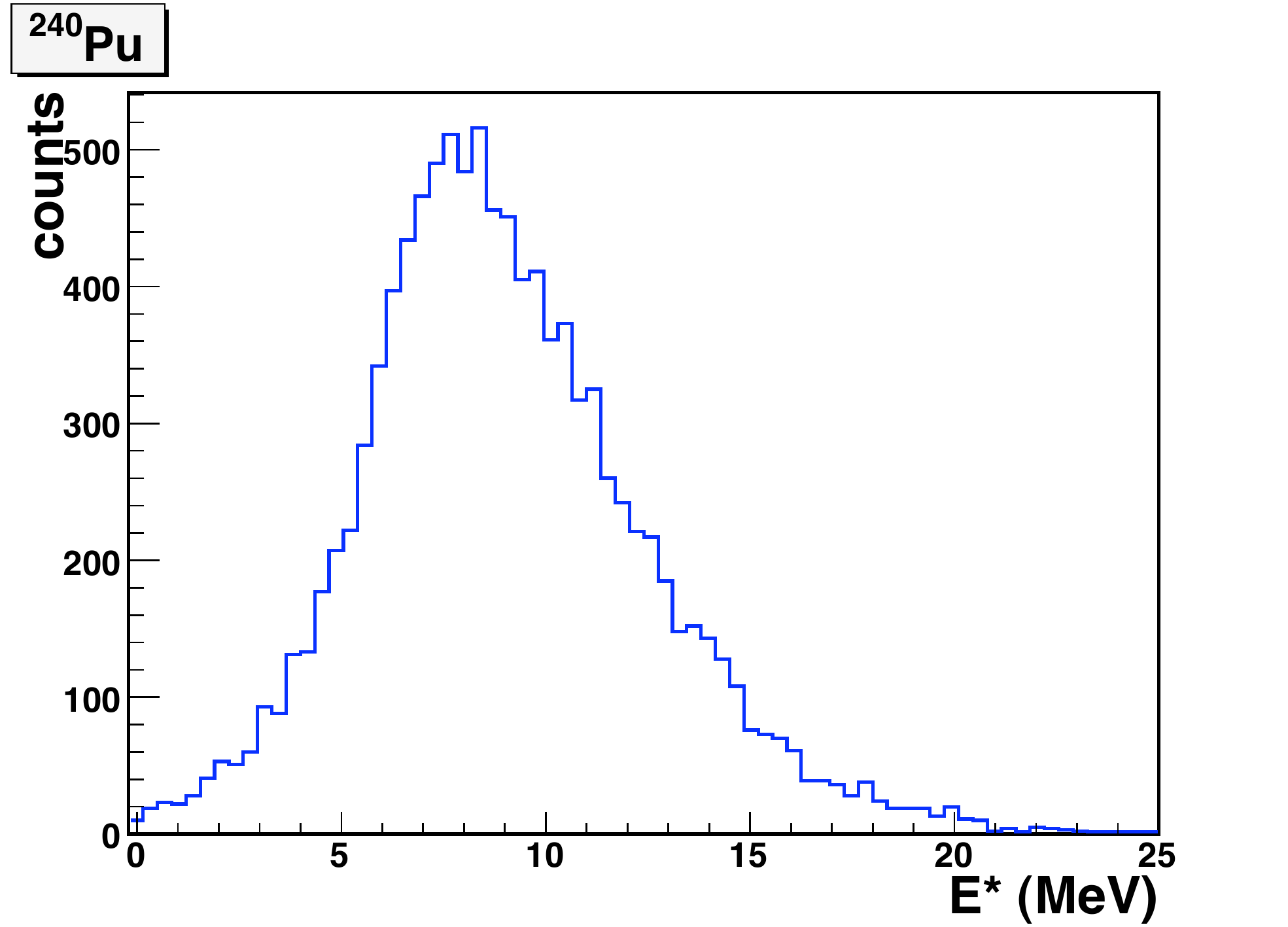}
\end{minipage}\hspace{2pc}%
\caption{\label{spider_eex}Excitation energy spectra associated to the B and Be  recoils in left and right panels, respectively.  }
\end{figure}

The different transfer channels are identified with an annular telescope, SPIDER, placed at the target position. It is composed of 2 double-sided silicon detectors, sectorised in polar and azimuthal angle $\Theta$ and $\Phi$, and covers angles from 35 to 53 $^\circ$. The multi-nucleon transfer cross section being centred at the grazing angle of 30 $^\circ$ in the laboratory reference frame, about 50 \% of the cross section is covered by the detector. In addition, its internal aperture is sufficient to let the fission fragments continue their flight without being intercepted. The resulting identification of the different transfer channels is reported in figure~\ref{spider_dEE}. Clearly, target recoils from C down to He are observed, corresponding to the formation of actinides from U to Cm. No transfer from $^{238}$U to the $^{12}$C target was observed in this experiment. Unfortunately, several technical problems that appeared during the experiment prevented from an isotopic resolution of the different target-like particles. A detail of the technical difficulties encountered and the procedure to correct for them is reported in~\cite{Derkx09, DerkxPhD}. For each proton transfer, the associated number of transferred neutrons is estimated from the ground-state Q value of the different channels~\cite{Karp}. For the three lighter transfers ({\it i.e.} heavier target) recoils, it is possible to estimate that C detection is associated to the production of $^{238}$U via elastic and inelastic  scattering and few percents of $^{237}$U. The B channels are mostly associated to $^{239}$Np and few percents of $^{240}$Np, and the Be channels are associated to 70 \% of $^{240}$Pu and 30 \% of $^{241}$Pu. Because the He and Li channels can be produced by different mechanisms, they are not considered further in the analysis. Indeed, He may be produced either in fusion-evaporation reaction, or in He transfer followed by the decay of $^8$Be, of which only one $^4$He is detected. Li can be the result of a Li transfer, or the decay of $^8$Be in two $^4$He, both being detected in the same detector pixel.   For each of the considered proton-transfer channels, the determination of the excitation energy is done considering the most probable associated neutron-transfer channel. Figure~\ref{spider_eex} shows the excitation energy for the B and Be channels, in coincidence with a fission fragment observed at the focal plane of VAMOS. Spectra associated to B and Be are characterised by a mean excitation energy of 7.5 and 9 MeV,  respectively, and both distributions show  a standard deviation of about 3.5 MeV. When no target recoil is measured in SPIDER, a fusion reaction is assumed.

\begin{figure}[h]
\begin{minipage}{16pc}
\includegraphics[width=14pc]{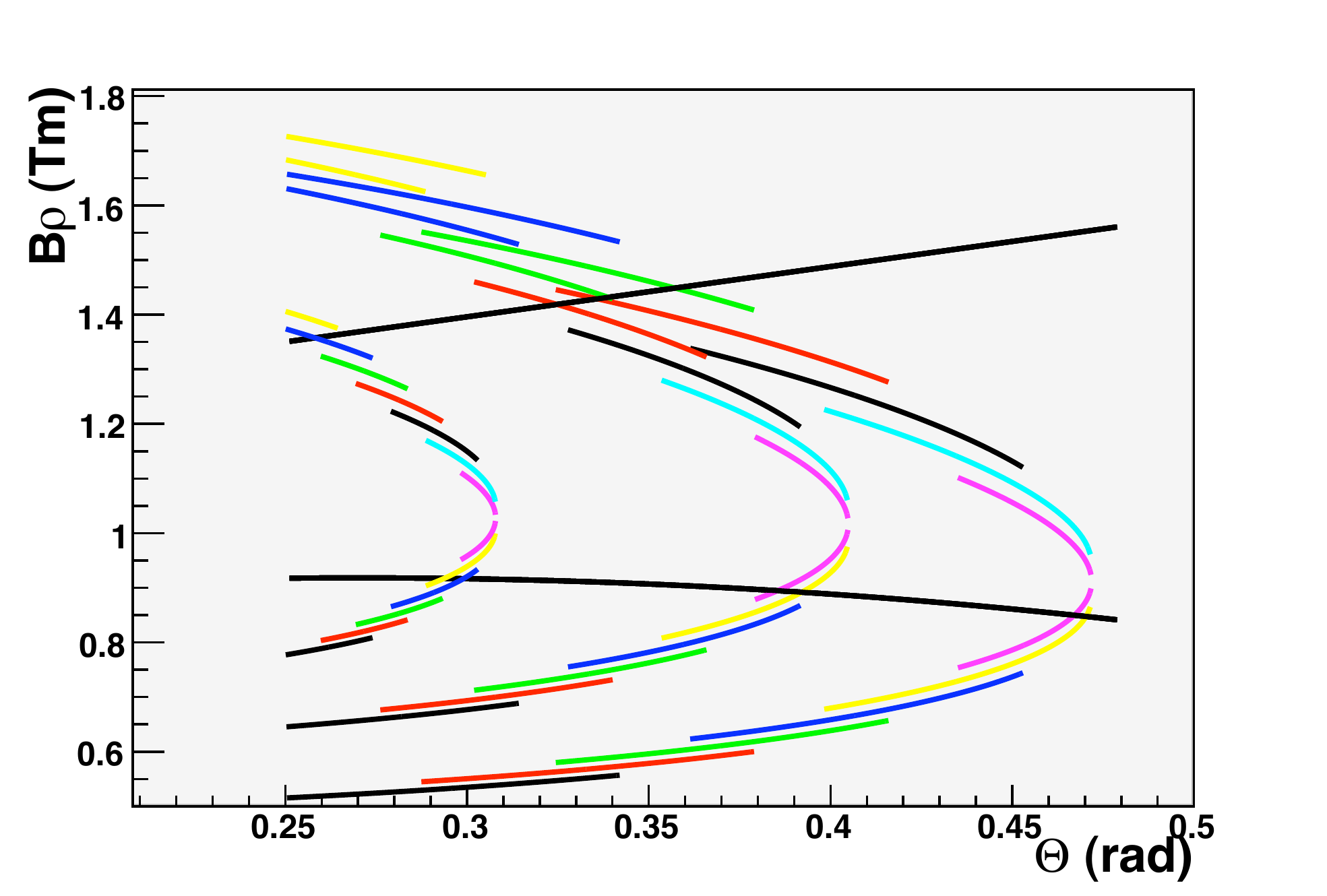}
\end{minipage}\hspace{2pc}%\vspace{-10pc}
\begin{minipage}{16pc}
\includegraphics[width=14pc]{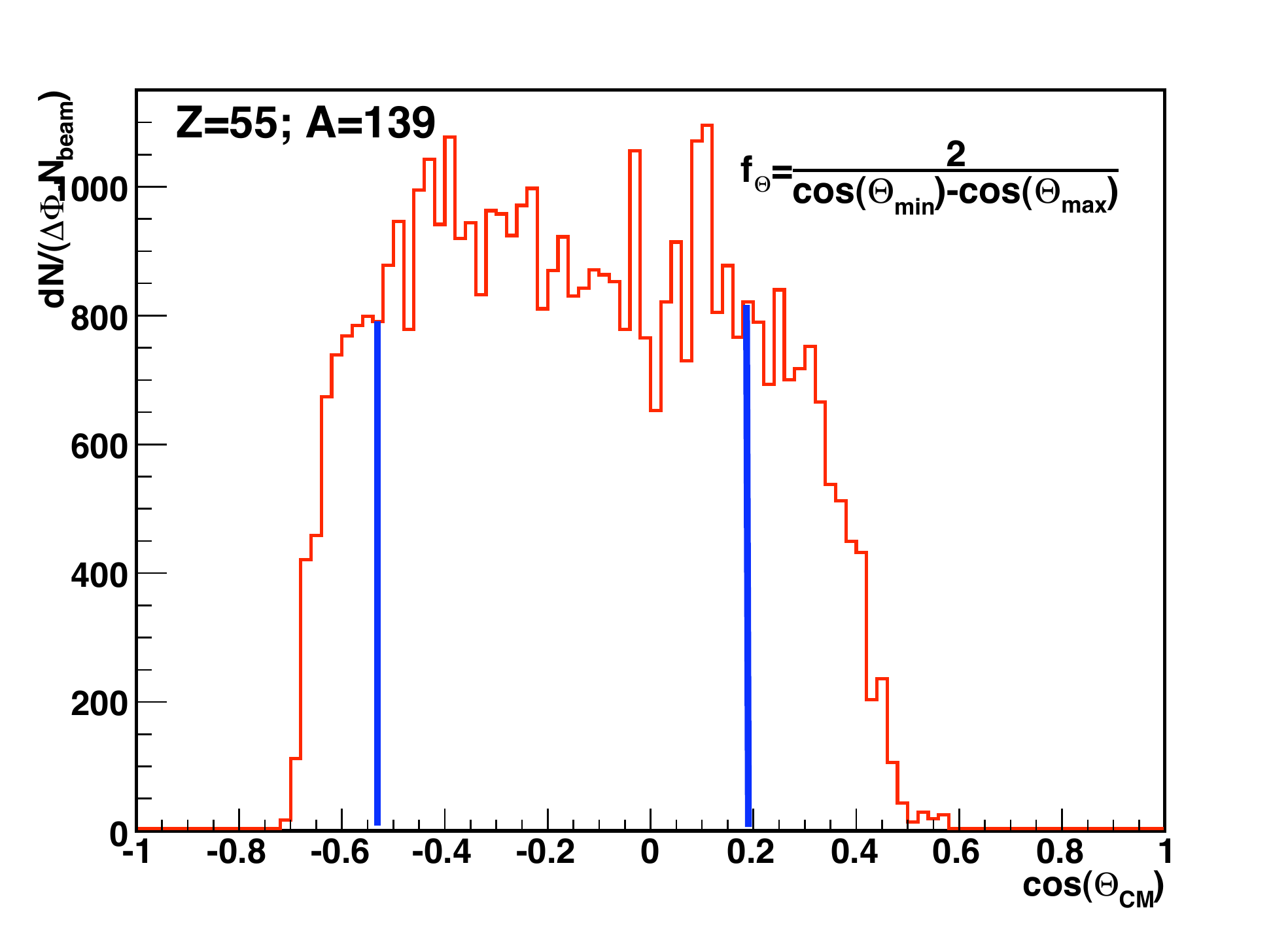}%\label{ang_dist}
\end{minipage}
\caption{\label{trans}Left: Scheme of the fission fragment kinematics in the laboratory reference frame. 
Each parabola corresponds to a different fission fragments with different fission velocity. The magnetic rigidity distribution is spread over about ten different ionic charge states. The momentum and angular cuts of the VAMOS spectrometer, covering the eight different values of the magnetic fields used, are shown by the two black diagonal lines. Right: Resulting angular distribution of $^{155}$Cs in the fissioning-nucleus reference-frame, summed over all ionic charge states. The blue lines indicate the angular aperture $\Theta_{min}$ and $\Theta_{max}$ of the spectrometer for this particular isotope transformed into the reference frame of the fissioning nucleus . The correction factor to take into account for the limited solid angle is indicated in the figure. }
\end{figure}

In the reference frame of the fissioning nucleus, each fragment has a velocity ({\it fission velocity}), which is characterised by the total kinetic energy defined by the Coulomb repulsion at scission, and the sharing between both fragments following the momentum conservation. This velocity is independent of the angle of emission, and a heavy fragment shows a smaller velocity than a lighter fragment.  In the reference frame of the laboratory, the fission-fragment velocity is correlated to the angle of emission in the reference frame of the fissioning nucleus. Each fragment has a velocity distribution, measured within the lower and upper angles of the VAMOS acceptance. In addition, the velocity distribution is spread over ten or so different charge states, leading to a broad magnetic rigidity distribution. These kinematics particularities are summarized in figure~\ref{trans}, left panel, where the magnetic rigidities as a function of the angle in the laboratory reference frame for 3 different fission fragments (one heavy, one average and one light) are sketched, as well as the limits of the angular acceptance of VAMOS. To obtain the isotopic yields of fission fragments, it is necessary to reconstruct the complete momentum of each isotope within the angular acceptance of VAMOS. For this purpose, eight different settings of the magnetic fields have been used.  Within these limits, shown in figure~\ref{trans}, left panel, each event is weighted by the total beam intensity and the azimuthal angular aperture of VAMOS, which is a function of the magnetic rigidity and the angle of the particle~\cite{Pulli}. With this correction applied, it is considered that within the angular cuts, the complete production intensity is measured. The method of yield reconstruction and transmission correction is described in details in~\cite{caamanotosubmit, delaunePhD}. 
The elastic scattering measured in SPIDER, which is tagged as a carbon event with no fission in coincidence is used to normalise the incoming beam intensity. A precision of 10\% is obtained in the normalisation. 

\begin{figure}[h]
\begin{center}
\begin{minipage}{18pc}
\hspace{-1.4pc}\includegraphics[width=20pc]{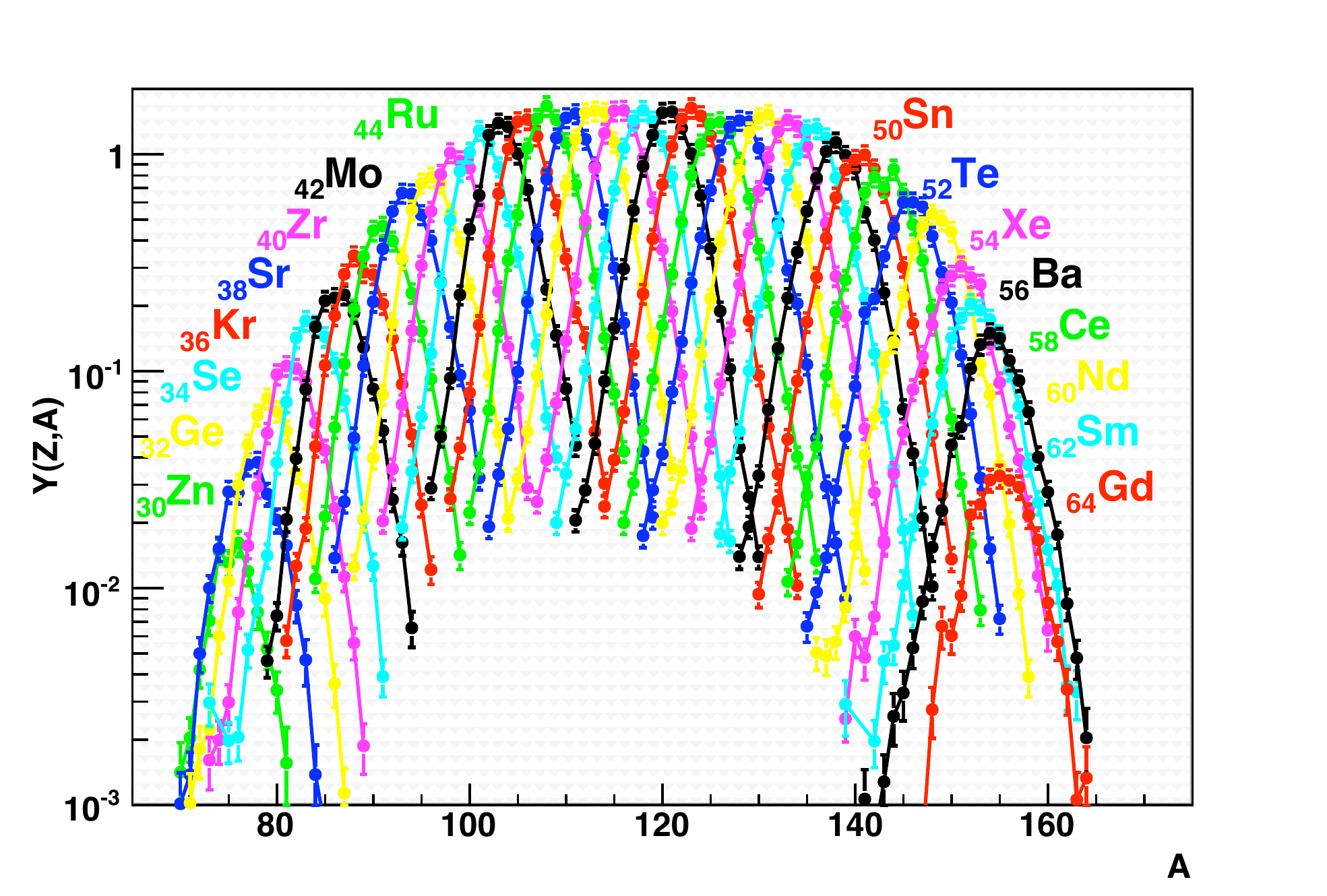}\label{YZA_Cf}
\end{minipage}
\end{center}
\begin{center}
\vspace{-1pc}
\begin{minipage}{15pc}
\includegraphics[width=14pc]{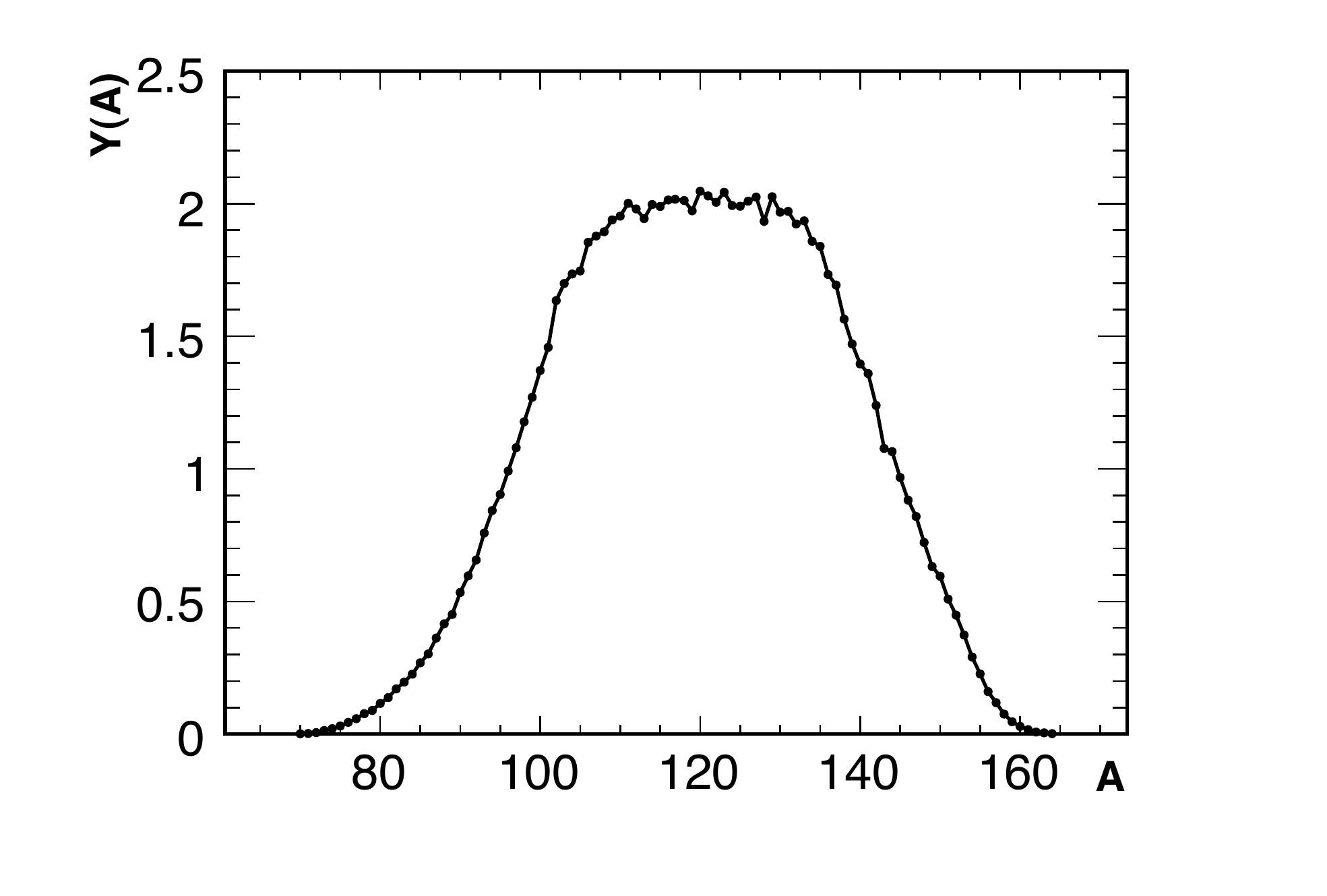}\label{YA_Cf}
\end{minipage}\hspace{-3.4pc}
\begin{minipage}{15pc}
\includegraphics[width=14pc]{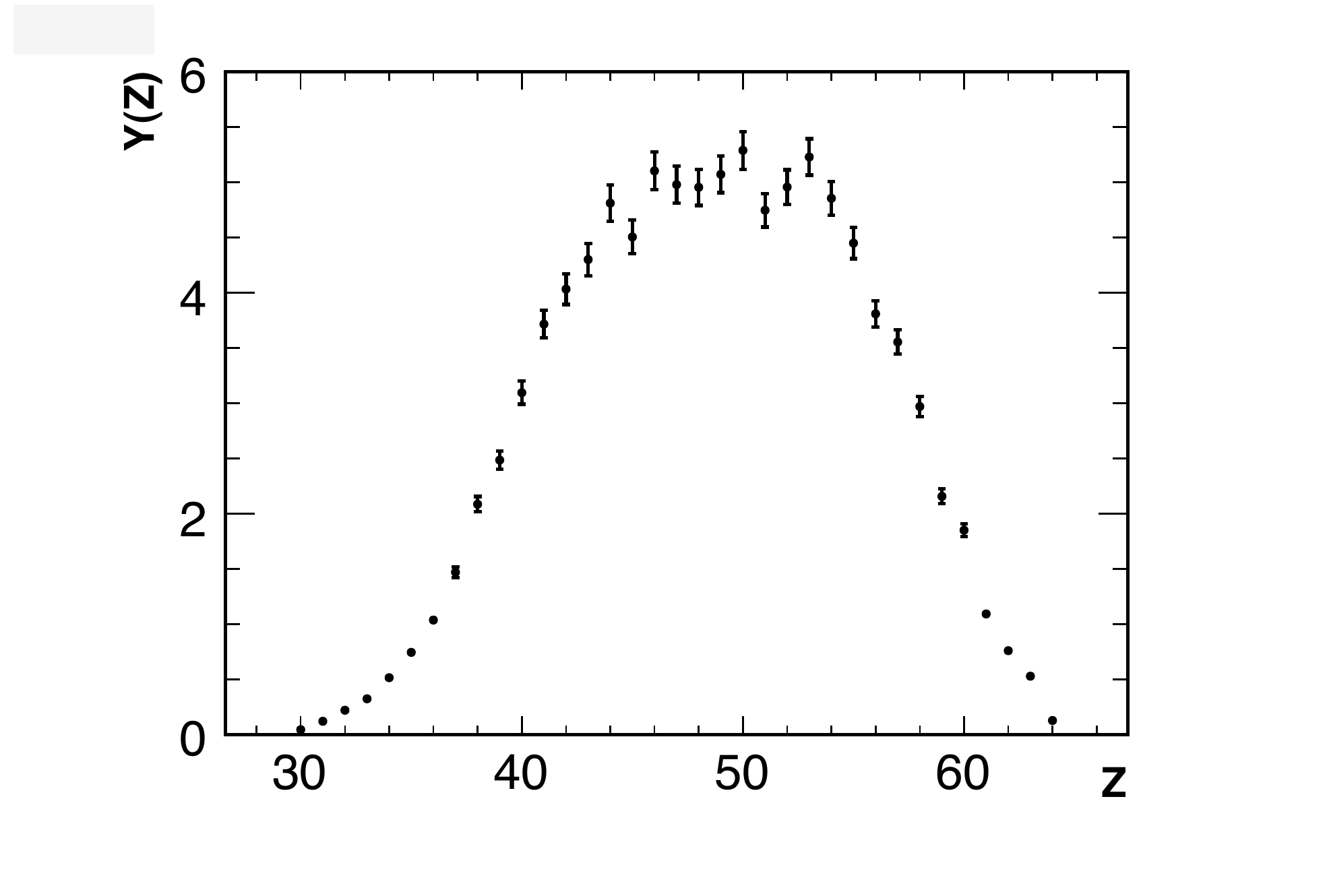}\label{YZ_Cf}
\end{minipage}
\end{center}

\caption{\label{Y_Cf} Upper panel: Isotopic distribution of fission fragments measured in the fusion-fission reaction $^{238}$U+$^{12}$C at 6.1 A MeV, from Z=30 to Z=63. Lower panel: The corresponding mass and atomic number distributions are displayed in the left and right panels, respectively. }
\end{figure}

The angular distribution of the fission fragments is then transformed into the reference frame of the fissioning nucleus, as well as the angular acceptance of the spectrometer, as sketched in figure~\ref{trans}, right panel. The isotopic yield is then defined as the integral of the angular distribution within the acceptance limits, corrected by a factor $f_\Theta$ that takes into account the limited solid angle sketched by the blue lines in figure~\ref{trans}, right panel. The limits of integration are different for each isotope. Among the 650 different isotopes measured, no sign of any anisotropy could be observed, and therefore the angular distribution is taken as flat and extended up to larger angles with the same amplitude. The isotopic yields measured with this procedure for fusion-fission reactions are displayed in figure~\ref{Y_Cf}, as well as the mass and atomic-number distributions. The element distribution shows a symmetric behaviour, with the yields of two complementary elements with respect to the atomic number of the fissioning nucleus ($Z=98$) being equal within the experimental uncertainties. This is a first confirmation that the method employed accurately corrects for the different transmission losses. The compound nucleus is produced with an excitation energy of 45 MeV, therefore the fission-fragment distributions are not expected to be influenced by shell-effects. The very asymmetric reaction $^{238}$U+$^{12}$C at Coulomb energy (10\% above the Coulomb barrier) leads to the production of a compound nucleus in rather clean conditions: quasi-fission is strongly hindered, and the angular momentum rather low (of the order of 25 $\hbar$). It is the first time that the isotopic distribution is measured in such reactions, and gives access to the influence of the charge-equilibration process on the evolution of the macroscopic part of the potential energy of the deforming nucleus.

\begin{figure}[h]
\hspace{1pc}
\begin{minipage}{15pc}
\includegraphics[width=16pc]{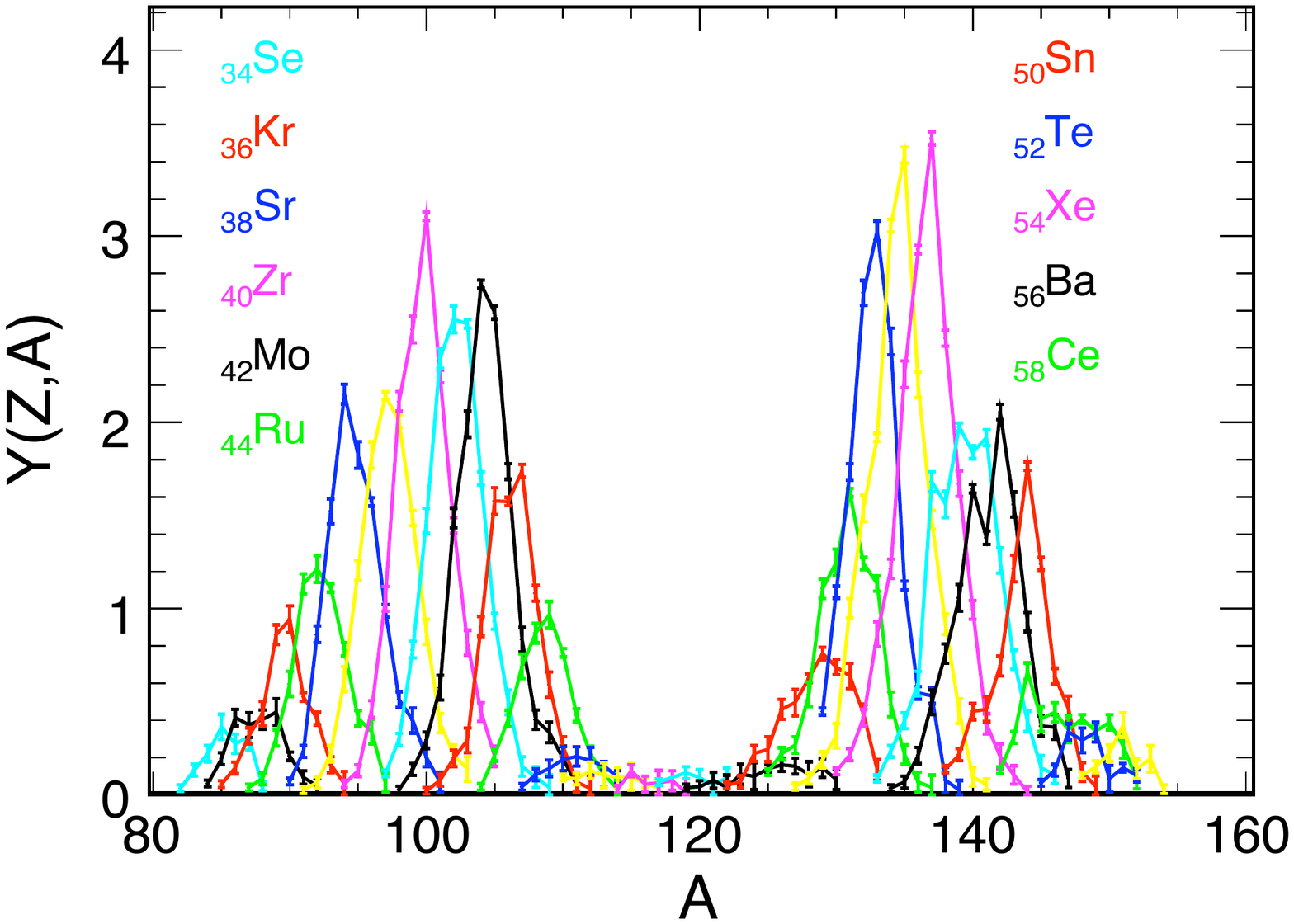}\label{YZA_Pu}
\end{minipage}\hspace{1pc}
\begin{minipage}{15pc}
\includegraphics[width=16.5pc]{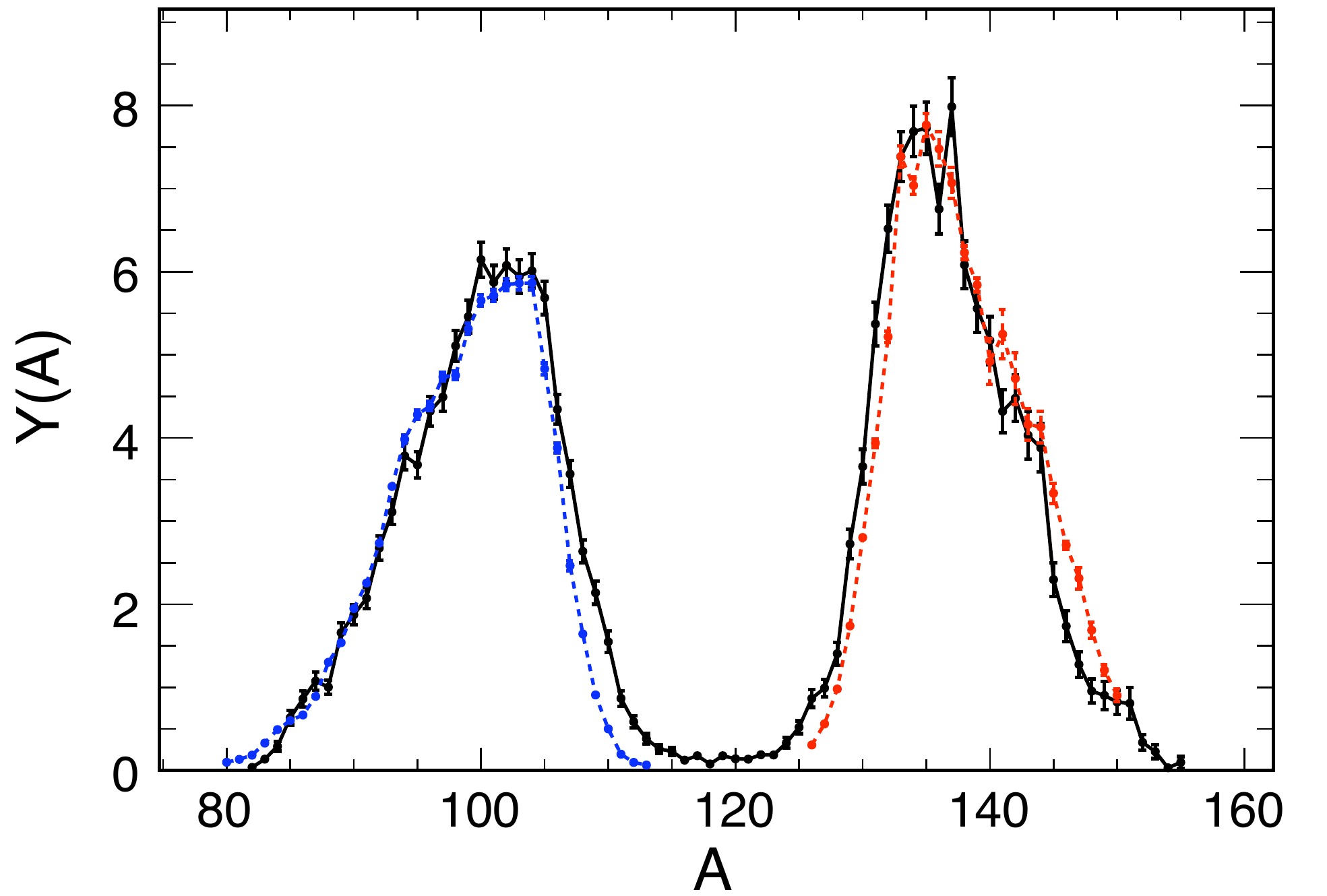}\label{YA_Cf}
\end{minipage}
\caption{\label{Y_Pu} Isotopic distribution of fission fragments measured in the two-proton transfer reaction $^{12}$C($^{238}$U, $^{240}$Pu)$^{10}$Be. The corresponding mass  distribution is displayed in the right panel, in black. Mass yields are compared to Lohengrin data obtained in normal kinematics, in blue~\cite{schmitt} and red~\cite{bail}. }
\end{figure}

The isotopic distributions of the fragments produced in transfer-induced fission are also accessible with the method presented above. Due to the technical difficulties encountered by the target-recoil detector SPIDER, only the two-proton transfer data are presented in this work, which correspond to the fission of a majority of $^{240}$Pu compound nuclei produced with an average excitation energy of 9 MeV, as shown in figure \ref{spider_eex}.  The isotopic distributions measured in the fission induced after two-proton transfer are displayed in figure~\ref{Y_Pu} for atomic number $Z$ ranging from 34 to 45 and 50 to 58. The corresponding mass yield distribution is compared to previous measurements obtained at the Lohengrin spectrometer using direct kinematics, and the resulting good agreement between different types of measures confirms the normalisation procedure and the spectrometer transmission estimation.  

\section{Fission experiment using inverse kinematics at higher bombarding energy}

With increasing bombarding energy, the compound nucleus resulting from fusion reactions is produced at higher excitation energy. Consequently, shell effects vanish, and a larger number of neutrons are evaporated by the system. The fission fragments are therefore sensitive to the evolution of the macroscopic part of the potential energy, and the different time-scales for neutron evaporation and deformation.  High-energy fusion-fission reactions in inverse kinematics may be an interesting technique to produce exotic-ion beams, as higher energy induces larger fusion cross section, the possibility to use thicker targets, and consequently higher yields of exotic nuclei, which can compensate the larger number of evaporated neutrons. In addition, high-energy fission allows to produce nuclei in the valley region, as well as on a broader range of elements~\cite{tarasovNIM, delaunecolloque}. GANIL offers the possibility to accelerate $^{238}$U beam up to 24 A MeV. This energy has been chosen to bombard 15 mg/cm$^2$ thick $^9$Be and $^{12}$C targets, leading to compound nuclei of $^{247}$Cm and $^{250}$Cf in the case of a fusion reaction, with an average excitation energy of 175 and 215 MeV, respectively.
In both reactions, the compound nuclei are very close ($^{247}$Cm and $^{250}$Cf) with a similar neutron  excess N/Z. For higher energy beams, a longer flight-path is requested, and therefore the LISE spectrometer~\cite{LISE} of GANIL has been chosen.  The identification of the fission fragments is based on the same technique as described previously, {\it i.e.} energy, energy-loss, magnetic rigidity and velocity measurements, which allow for identifying each isotope in atomic number $Z$, mass number $A$, and ionic charge state $q$. Four different settings of the spectrometer magnetic rigidity have been used, 1.8, 1.9, 2, and 2.1 Tm. This spread in magnetic rigidity span over the complete magnetic rigidity distribution. Due to the small magnetic rigidity acceptance of the spectrometer  ($\pm$ 2.5 \%), the different settings do not overlap in momentum distribution.  However, the small acceptance of the spectrometer induces square acceptances in angle and magnetic rigidity, which can therefore be computed easily with kinematics simulation to recover the initial yield production.

\section{Results and discussion} 
\begin{figure}[h]
\hspace{-1pc}
\begin{minipage}{15pc}
\includegraphics[width=15pc]{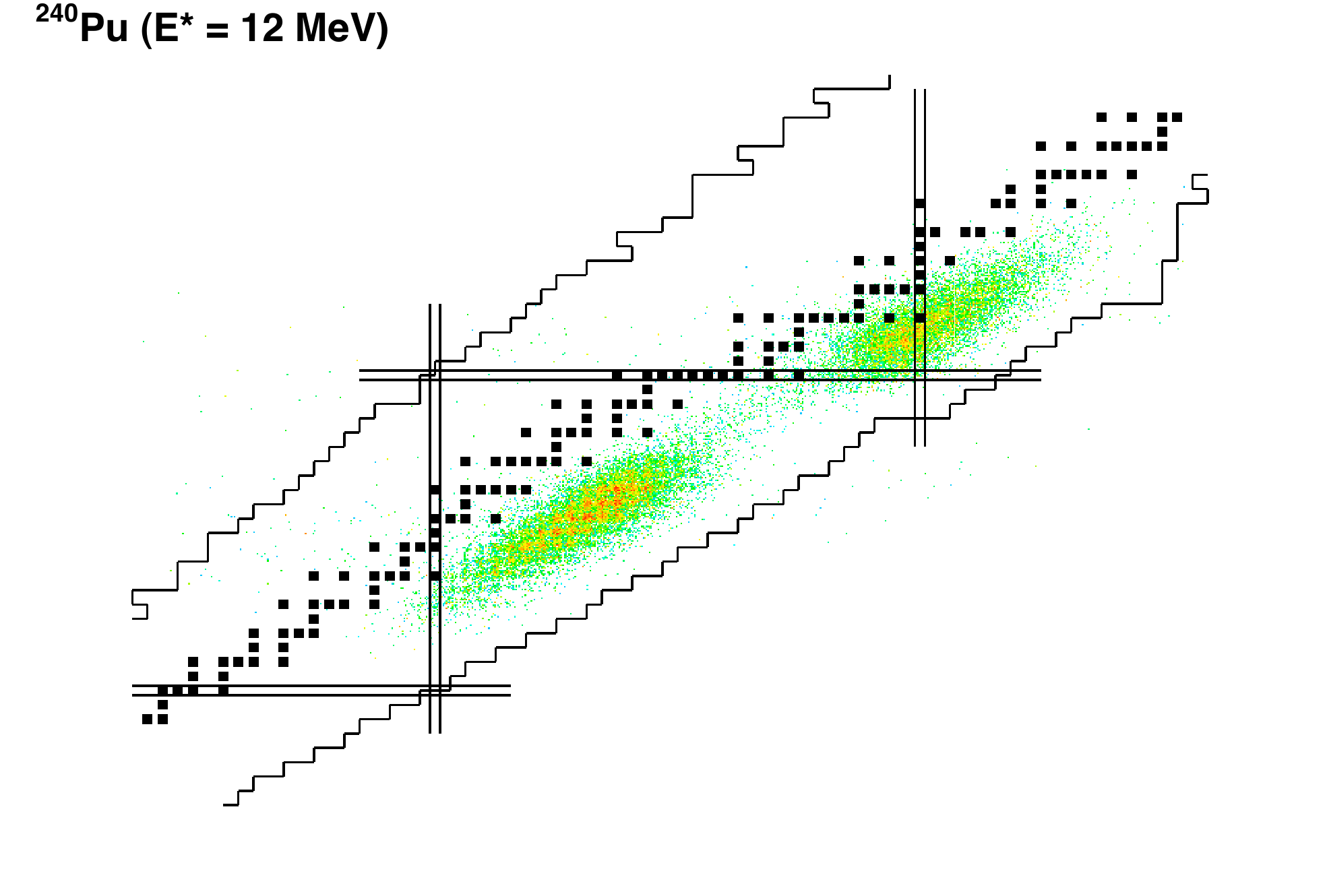}
\end{minipage}\hspace{-4pc}
\begin{minipage}{15pc}
\includegraphics[width=15pc]{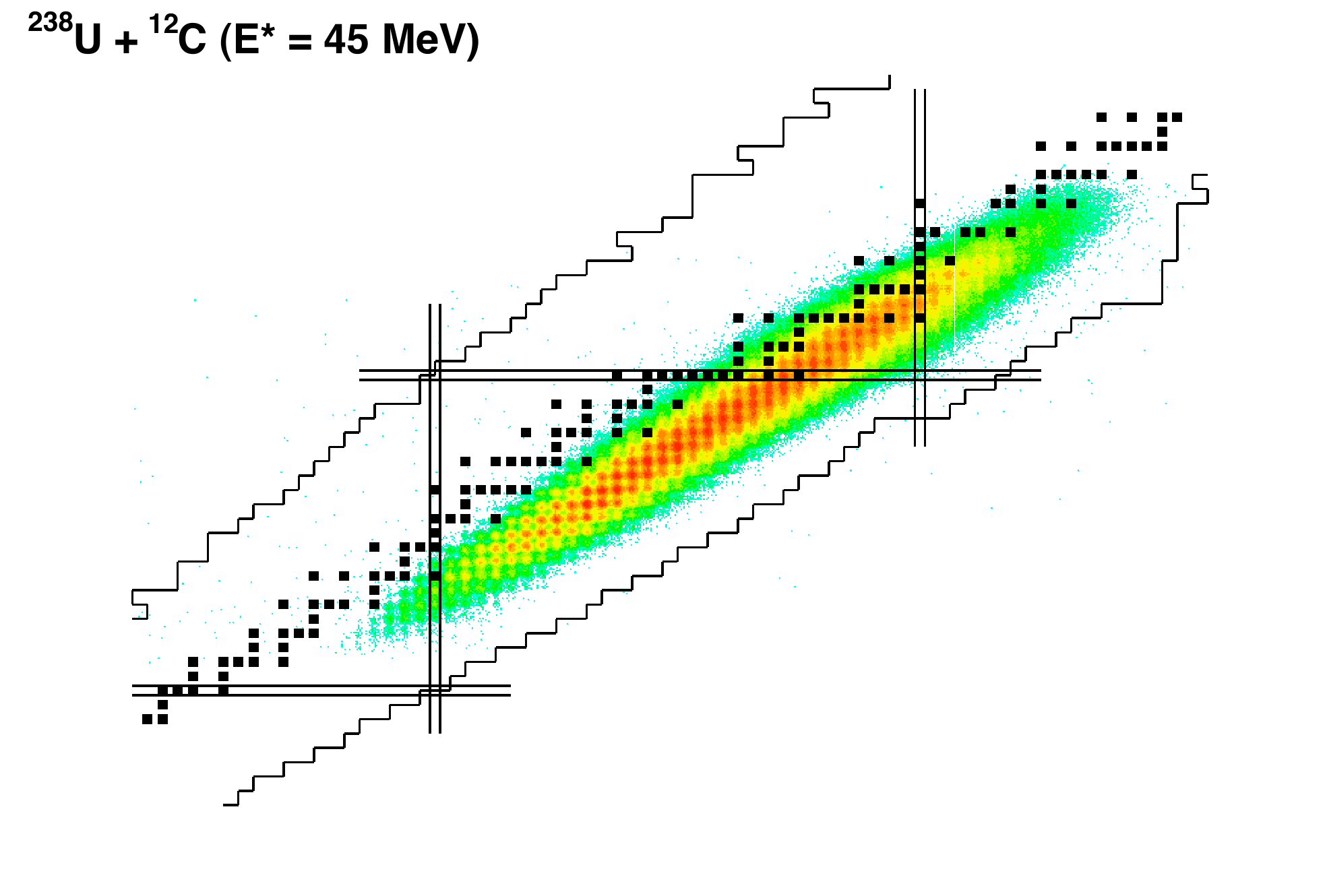}
\end{minipage}\hspace{-4pc}
\begin{minipage}{15pc}
\includegraphics[width=15pc]{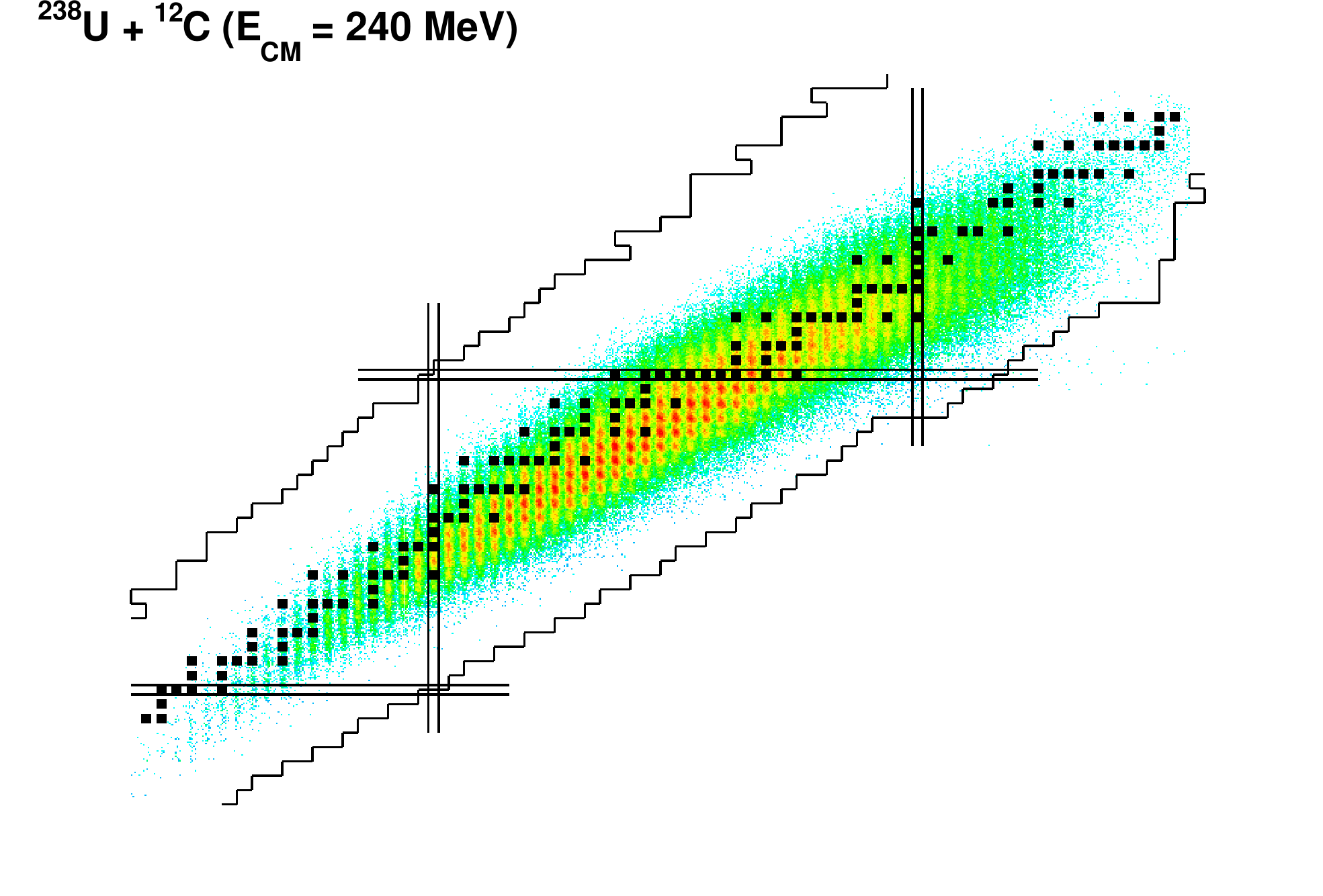}
\end{minipage}
\caption{\label{cartes} Isotopic distributions of fission fragments measured in three different reactions, displayed on a nuclide chart. The left panel corresponds to the production in low-energy fission induced in two-proton transfer (see text for details). The middle panel corresponds to fusion-induced fission, $^{238}$U beam on $^{12}$C target, with the production of a compound nucleus of $^{250}$Cf  at an excitation energy of 45 MeV. The right panel corresponds to  fission-fragment production in the same reaction but at higher bombarding energy, with an energy in the centre of mass of 240 MeV.}
\end{figure}

The resulting isotopic distributions measured in fission induced in two-proton transfer and fusion on $^{12}$C at 6 and 24 A MeV bombarding energy  are presented in three different nuclide charts in figure~\ref{cartes}. Several hundreds of isotopes are measured in each system over 2 orders of magnitude, except for the transfer-induced fission where the lack of statistics restricts the sensitivity to one order of magnitude. 

\begin{figure}[h]
%\begin{minipage}{14pc}
\includegraphics[width=30pc]{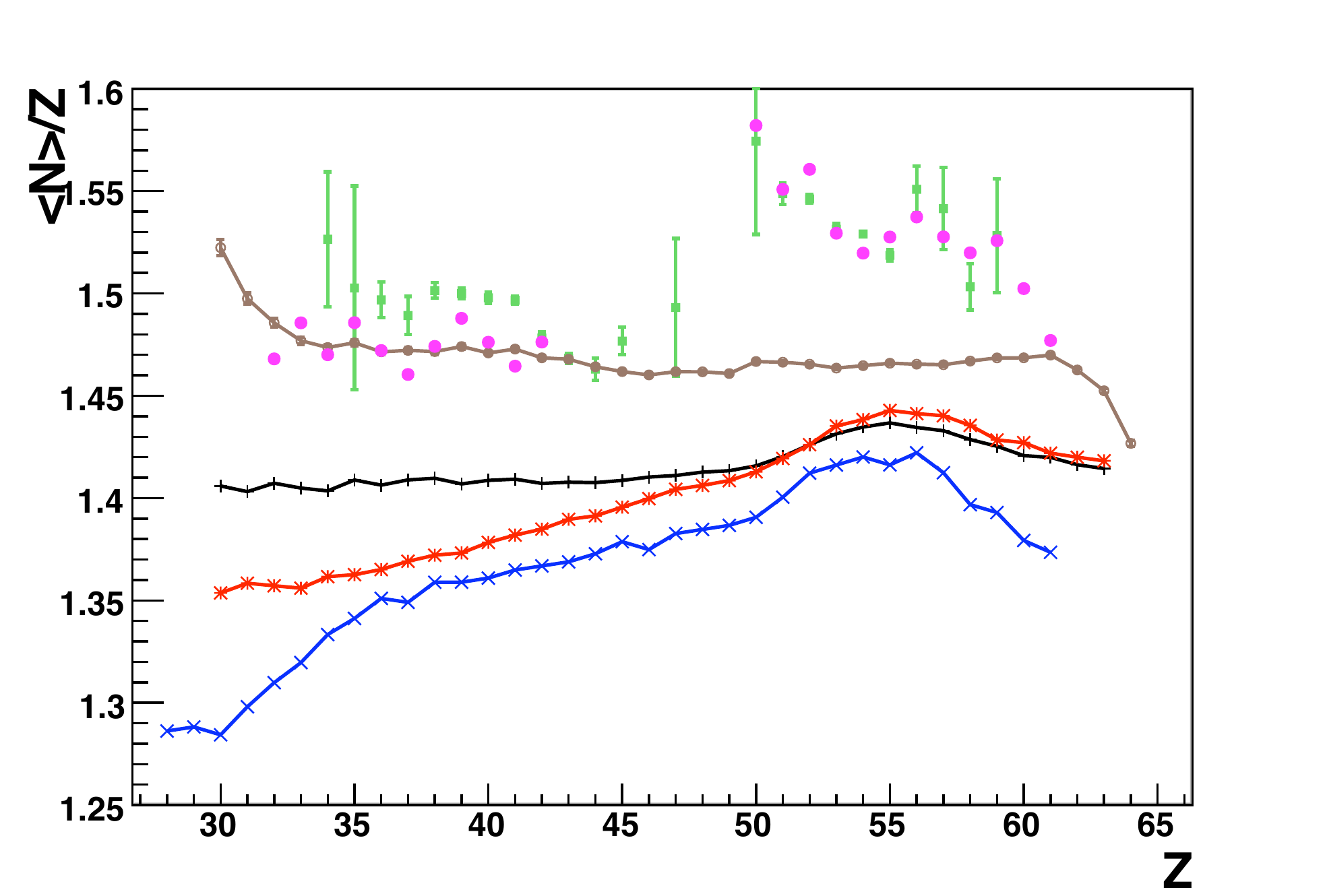}
\caption{\label{NoZ}Neutron excess $<$N$>$/Z of the fission fragments for the different fission reactions investigated in this work: two-proton transfer-induced fission (green full squares), fission of the compound nucleus $^{250}$Cf at 45 MeV of excitation energy (brown full circles), fission induced at higher bombarding energy, leading to the compound nucleus $^{247}$Cm with 175 MeV of excitation energy (black crosses), and  $^{250}$Cf with 215 MeV (red asterisks). In addition, data from spallation-induced data where electromagnetic-induced fission (purple full circles) could be separated from fragmentation-induced fission~\cite{pereira} (blue crosses) are displayed for comparison.}
%\end{minipage}\hspace{2pc}%
\end{figure}

In order to determine the gross properties of the fission fragments,  the neutron excess is defined as the average neutron number $<$N$>$ of an isotopic distribution, divided by the corresponding atomic number $Z$. The evolution of the neutron excess of the fission-fragment distributions measured in the three systems referred to in figure~\ref{cartes} are displayed in figure~\ref{NoZ}. In addition, the systems $^{247}$Cm produced in fusion reaction at 175 MeV of excitation energy, and fission fragments produced in fragmentation and electromagnetic-induced reactions~\cite{pereira} are displayed for comparison. 

The neutron-excess of fission fragments produced in low-energy fission (transfer-induced fission) are similar to electromagnetic-induced fission of $^{238}$U~\cite{pereira}. In both reactions, characterised by an asymmetric fragment distribution, the neutron excess shows a step behaviour, with two different values for the light and the heavy fragments.  Heavy fission fragments are produced with larger neutron excess than light fragments. This property is known from neutron-induced fission as charge polarisation  and is interpreted as the result of the influence of the symmetry energy of the macroscopic part of the nucleus potential energy, enhanced by the influence of shell effects. The shell structure in the heavy nascent fragment induces a larger neutron excess, while the neutron excess of the light fragments reflect the complementary part  of the fissioning nucleus. In addition to the number of neutrons and protons defined at their formation, the shape of the fragments at scission impacts on their neutron excess; after their separation, the fragments produced in almost spherical shape evaporate few or no neutrons compared to those produced in deformed shapes. Indeed, a very large neutron excess of 1.57 is observed for Sn. 

The neutron excess observed in fusion-induced fission reactions, leading to the compound nucleus $^{250}$Cf at an excitation energy of 45 MeV,  shows a constant value of 1.47, over the complete element distribution. The excitation energy of the compound nucleus is consistent with the evaporation of about 6 neutrons. These 6 neutrons may be evaporated before the saddle point, during the elongation from saddle to scission, or after the scission point, depending on the time needed by the fission process compared to the time for neutron evaporation.   

Considering the reaction leading to the compound nucleus $^{247}$Cm with an average excitation energy of 175 MeV, the overall neutron excess decreases, in agreement with a higher excitation energy induced in the reaction. However, a charge polarisation is also observed as a step structure, revealing the influence of low-energy fission in the fission fragment distribution properties. Indeed, in the fission at low excitation energy, the compound nucleus  produces heavy fragments in the $Z=54$ region with a larger neutron excess. Therefore, the region around Z=54 is especially strongly populated by low-energy fission, while higher excitation energy fission leads to fragments with smaller neutron excess, in the region around $Z=48$. Thus, the bump observed close to $Z=54$ in the neutron excess distribution reveals that in a significant proportion of the fission decays, a large part of the excitation energy may be released before the nucleus reaches the saddle deformation, and that therefore, the fission-path in the potential energy landscape is influenced by shell effects that appear at lower excitation energy. This behavior contrasts with the lower excitation energy fusion-fission data, where no sign of shell structure is observed in the neutron excess distribution. The contradictory observation that a more excited compound nucleus is more sensitive to the shell structure can find  enlightenment in the deformation dynamics. If at higher excitation energy the time needed to reach the saddle deformation is longer, the nucleus has additional time for neutron evaporation, which cools down the fissioning nucleus, which then is sensitive to the shell structure during the emergence of the two fission fragments. 
The neutron excess of the light fragments of $^{247}$Cm is compatible with a long evaporation chain of 15 neutrons. A longer deformation time at high excitation energy may be possibly interpreted  as resulting from the increase of the dissipation with excitation energy, as suggested in~\cite{hofman}. A possible effect of transient time, resulting in a fission delay at excitation energies above 100 MeV may also be inferred~\cite{jurado}. A possible influence of the angular momentum on the dissipation has recently been reported~\cite{Ye}, which is also in favour of a different dissipation regime between the different reactions, where angular momentum is varying from about 25 to 80$\hbar$. Further investigations based on de-excitation simulation should enable to validate the different hypothesis. 

When the excitation energy of the compound nucleus is still increasing, as for $^{250}$Cf at 215 MeV of excitation energy,  the overall neutron excess still decreases, but the charge polarisation remains observable, showing that low excitation-energy fission is still influencing the fission-fragment distributions. However, if the bump around $Z=54$ is still present , an additional feature arises as the neutron excess of the light fragments varies with the atomic number of the fragments.  As may be seen in figure~\ref{NoZ}, a slope is observed, the neutron excess is increasing with the atomic number of the fragments. A slope is also observed in spallation reaction, and can be understood as an experimental signature for the contribution of different fissioning systems. Indeed, in spallation reactions, the intra-nuclear cascade inside the nucleus leads to a wide variety of fissioning nuclei, produced with a broad distribution of excitation energy. The different nuclei show different neutron excess, while their fission may be influenced by shell structure in the case the excitation energy is sufficiently low at the saddle deformation. The neutron excess of the complementary light fragments reflects the neutron excess of their mother nucleus. The lighter fissioning systems are produced with smaller neutron excess, and produce lighter complementary fragments, which results in a slope in the neutron excess~\cite{lukic}. If the slope in neutron excess is present in the  $^{250}$Cf at 215 MeV of excitation energy, it reveals that between the two fusion-reactions studied in the present work, the regime of reaction is much different. Either the slightly increased energy in the centre of mass (240 compared to 190 MeV), or the slightly increased target mass-number, induce a growing contribution from incomplete fusion, and a larger emission of pre-equilibrium particles. This leads to the formation of an ensemble of fissioning systems, varying in Z, A, produced over an excitation energy distribution.  If they fission at low excitation energy, the compound nuclei  produce heavy fragments in the same $Z=54$ region with larger neutron excess, while fission at higher excitation energy produces symmetric fission and peaks at lower Z. If the compound nucleus is even lighter, the fission fragments are lighter, too, as in the case of spallation. 

%This feature reveals that at higher energy, the initial kinetic energy is not damped completely into intrinsic excitation energy in the formation of the compound nucleus, but that some of the initial kinetic energy is evacuated of the system via pre-equilibrium emission. Consequently, the fissioning system is not defined in (Z,A), and the excitation energy of the compound nucleus may expand over a broad range. The fission-fragment is therefore still influenced by low-excitation energy fission, which is known to stabilise the heavy fragment bump. The light fragments are then a complementary part of the fissioning system. As lighter fissioning systems arise, they are also less neutron-rich, as the proton emission will be associated to neutron evaporation, and the lighter fissioning systems fission with lighter fragments. Therefore the neutron excess of the light fragments will feel the neutron excess of the fissioning nucleus, while the heavy fragments are more influenced by shell structure. This explains the slope.

\section{Post-scission neutron multiplicities}
\begin{figure}[h]
%\begin{minipage}{14pc}
\includegraphics[width=30pc]{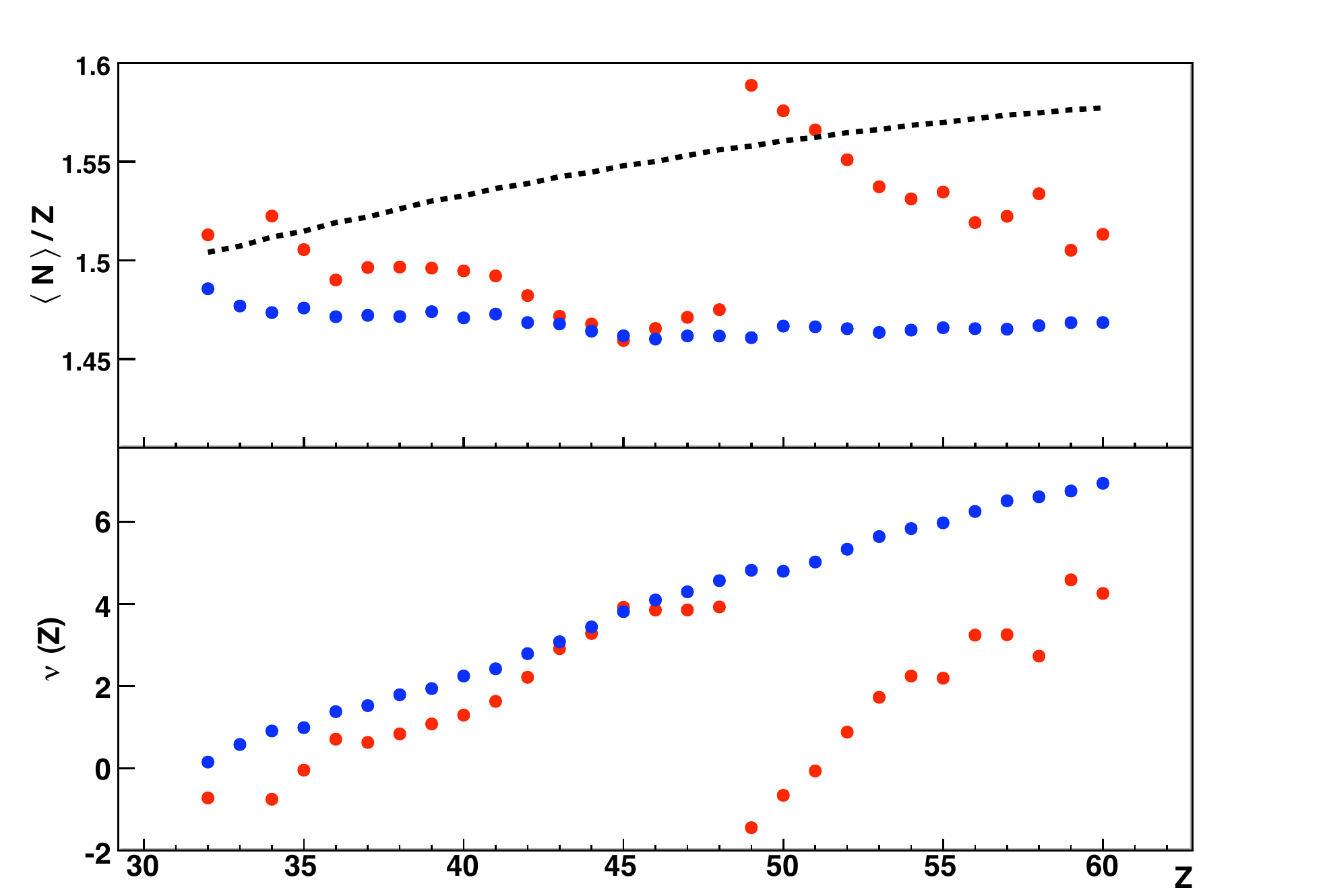}
\caption{\label{nu}Upper panel: Neutron excess $<$N$>$/Z as expected from the scission-point model (dashed line). It is compared to the neutron excess of fission fragments measured in 2-proton transfer induced fission (red full circles),  and fusion-induced fission (blue full circles) at 45 MeV of excitation energy. 
Lower panel: the difference between the measured neutron excess and the expected one at scission deformation leads to an estimation of the post-scission neutron multiplicity (same color code).}
%\end{minipage}\hspace{2pc}%
\end{figure}

Considering the scission-point model of Wilkins {\it et al.}, it is possible to determine the neutron excess of the fission fragments that minimize the macroscopic potential energy surface (no shell influence is considered). Despite the lack of dynamics effects in this model, it is the only one that allows for varying the atomic and neutron number independently, all other models being based on the unchanged charge density hypothesis. The resulting neutron excess of the most probable isotope is shown in figure~\ref{nu} as a dashed line. The observed curvature is similar to the curvature of the valley of stability: the heavier fragments are more neutron rich in order to compensate for increasing Coulomb energy. For comparison, the measured neutron-excess for the two compound nuclei $^{240}$Pu (two-proton transfer) and $^{250}$Cf at 45 MeV are displayed. The difference between the dashed line and the experimental points comes from the post-scission neutron evaporation, as well as a possible lack of the description, which is limited to the macroscopic part of the potential only. This difference is plotted in the lower panel of figure~\ref{nu}. In low-energy fission, the well-known saw-tooth behaviour of the neutron yield is observed, reflecting the shell influence on the scission configuration~\cite{Wilkins}, whereas a steady increase of the neutron multiplicity with the atomic number of the fragments is observed at higher excitation energy. This steady increase was already observed in direct neutron measurement in similar systems~\cite{hinde}. 

\section{Conclusion}
A new technique is reported for investigating the fission-fragment properties. The technique is based on inverse kinematics to produce fission-fragments in-flight. They are then identified  with different spectrometers of GANIL. This technique allows for the identification of the complete isotopic distributions of the fission fragments. The different nuclear reactions  investigated give access to fissioning systems at different excitation energies. The resulting impact of the initial conditions on the final fission-fragment distributions are discussed. 
The influence of shell structure, its excitation-energy fade, as well as its persistence depending on the deformation dynamics are being perceived with new type of experimental data. 
The isotopic distribution of fission fragments gives a severe frame to enlighten the influence of the different properties of nuclear matter that play a role in the fission process.

\smallskip

\end{document}